\documentclass[useAMS,usenatbib]{mn2e}
\pdfoutput=1
\usepackage[dvips]{graphicx}
\usepackage{dblfloatfix}
\usepackage{amsmath, amssymb}
\usepackage[breaklinks=true,colorlinks=true,linkcolor=blue,
urlcolor=black,citecolor=black,bookmarks=true,bookmarksopenlevel=2]{hyperref}
\usepackage{aas_macros}
\usepackage{import}
\usepackage{enumitem}
\usepackage{float}
\newcommand{\D}[1]{\ensuremath{\text{d}#1}}

\newcommand{\B}[1]{\bmath{#1}}

\title[The Anisotropic Line Correlation Function]{The Anisotropic Line Correlation Function as a Probe of Anisotropies in Galaxy Surveys}
\author[A. Eggemeier et al.]{A. Eggemeier$^{1,2}$\thanks{a.eggemeier@sussex.ac.uk}, T. Battefeld$^{1,3}$\thanks{tbattefe@gmail.com},
  R. E. Smith$^{2}$\thanks{r.e.smith@sussex.ac.uk} and J. Niemeyer$^{1}$\thanks{niemeyer@astro.physik.uni-goettingen.de}\\
  $^{1}$Institut for Astrophysics, University of G\"ottingen,
  Friedrich-Hund Platz 1, D-37077, Germany\\
  $^{2}$Astronomy Centre, University of Sussex, Falmer, Brighton, BN1 9QH, UK\\
  $^{3}$Department of Physics, University of Minnesota,
1023 University Dr., Duluth, MN 55812, U.S.A.\\
}
\begin{document}

\date{}


\maketitle

\label{firstpage}

\begin{abstract}
  We propose an anisotropic generalisation of the line correlation function (ALCF) to separate and quantify phase information in the large-scale structure of galaxies. The
  line correlation function probes the strictly non-linear regime of structure formation and since phase information drops out of the power spectrum, the line correlation
  function provides a complementary tool to commonly used techniques based on two-point statistics. Furthermore, it is independent of linear bias as well as the Gaussian
  variance on the modulus of the density field and thus may also prove to be advantageous compared to the bispectrum or similar higher-order statistics for certain cases. For
  future applications it is vital, though, to be able to account for observational effects that cause anisotropies in the distribution of galaxies. Based on a number of
  numerical studies, we find that our ALCF is well suited to accomplish this task and we demonstrate how the Alcock-Paczy\'{n}ski effect and kinematical redshift-space
  distortions can in principle be measured via the ALCF.
\end{abstract}



\section{Introduction}
\label{sec:introduction}

Galaxy surveys map a growing fraction of our Universe at ever increasing accuracy and are routinely used to test cosmological models. So far, agreement with the $\Lambda$CDM
model, a spatially flat universe primarily filled with dark matter and dark energy in the form of a cosmological constant, is excellent.  Among their key science questions are
the nature of dark matter, dark energy, and the concrete model of the early Universe during which primordial density fluctuations were generated. A range of next generation
experiments, such as DESI \citep{desi}, Euclid \citep{laureijs2011} and SKA \citep{dwedney2009}, etc. are designed to shed light onto these questions.

To make progress, it is crucial to optimize the amount of information extracted from the raw data that these experiments produce. Up until recently, the majority of the
literature focused on two-point statistics, i.e.~ the power spectrum $P$ or the two-point function $\xi$, to analyse the observed distribution of galaxies. If the density
field were Gaussian, two-point statistics would be sufficient to specify the statistical properties of matter fluctuations.  However, during the non-linear stage of
gravitational interaction, where structures like filaments, clusters and eventually galaxies form, any initial density field becomes highly non-Gaussian. Hence, by focussing
on two-point statistics, a considerable part of the available information is ignored.

Consequently, new statistical measures have received significant attention. Besides computing higher-order correlations \citep[e.g.][]{Peebles1975, peebles, Frieman1994,
  Scoccimarro1998}, it was suggested to employ genus statistics \citep{Gott1986, Hoyle2002, Hikage2002}, Minkowski functionals \citep{Mecke1994, Hikage2003} and
others. However, all of these measures suffer from the conceptual limitation that they are correlated with the two-point function. Thus, an intriguing idea is to exclusively
analyse information not already contained in two-point statistics, i.e.~\emph{phase information}.

Correlations in the phases of Fourier coefficients emerge as a consequence of non-Gaussianity; since two-point statistics depend on the amplitudes of Fourier coefficients
only, they are blind to phase information. The phase factors of the matter density field have been studied many years ago, but much of the original work in \cite{Ryden:1991fi,
  1992ApJ...396..379S, Jain:1998it} focused on the evolution of single phases away from their initial values. A first statistic that described the phase difference between
neighbouring Fourier modes was presented in \cite{Coles:2000zg, Chiang:2000vt, Watts:2002py}. Shortly thereafter, their considerations were generalised by
\cite{Matsubara:2003te}, who quantified phase information by computing the joint probability density of phase factors, which illuminated its relation to higher-order spectra
(see also \citealt{Hikage2004, Matsubara:2006ep}). In \cite{Hikage:2005ia}, the probability densities were applied to a galaxy catalogue from the SDSS DR-2, giving constraints
on bias models. Recently, a new measure for phase information, the line correlation function, was introduced by \cite{linecorr} as a spherically-averaged three-point
correlation of phase factors. In the follow-up paper \citep{Wolstenhulme:2014cla} it was computed perturbatively and related to the bispectrum and power spectrum at leading
order.

The aim of this paper is to extend the line correlation function to account for line-of-sight and transverse distance scales separately. The introduction of such an
anisotropic line correlation function opens up the possibility to detect anisotropies and distortions along the line-of-sight, which are inherently present in any galaxy
survey due to redshift-space distortions.

The paper is organised as follows: after a brief review of the definition and properties of the line correlation function in Sec.~\ref{sec:ILCF}, we present and test a
modification in Sec.~\ref{sec:ALCF}, which is capable of quantifying anisotropies. In Sec.~\ref{sec:RSD} this new formalism is applied to simple mock fields and we investigate
the sensitivity to the Alcock-Paczy\'{n}ski effect and kinematical redshift-space distortions in comparison to results based on the two-point function. We conclude in
Sec.~\ref{sec:conclusion}.


\section{The Isotropic Line Correlation Function}
\label{sec:ILCF}

In this section, we review the isotropic line correlation function (ILCF), which serves as the starting point for our generalisation to the anisotropic line correlation
function (ALCF) in Sec.~\ref{sec:ALCF}.

\subsection{Definition}
\label{sec:ILCF.definition}

Given a density field $\delta(\B{x})$, the ILCF measures correlations in its Fourier phase factors which are obtained by dividing amplitudes in Fourier space\footnote{The zero
  mode is conventionally set to zero, i.e.~$\epsilon_{\B{k}=0} = 0$.},
\begin{align}
  \delta(\B{x}) \hspace{0.5em}\overset{\text{FT}}{\longrightarrow}\hspace{0.5em} \delta_{\B{k}}
  \hspace{0.5em}\overset{\text{whitening}}{\longrightarrow}\hspace{0.5em} \epsilon_{\B{k}} \equiv \frac{\delta_{\B{k}}}{|\delta_{\B{k}}|}
  \hspace{0.5em}\overset{\text{IFT}}{\longrightarrow}\hspace{0.5em} \epsilon(\B{x})\,.
  \label{eq:whitening}
\end{align}
The \emph{whitened} density field $\epsilon(\B{x})$ is thus devoid of any information probed by two-point statistics. Consequently, the simplest measure of phase information
must be based on the three-point correlator and since the whitening transformation tends to collapse elongated objects to thin line segments, as was observed in
\cite{linecorr}, these three points are chosen to be distributed equidistantly on a straight line. More precisely, and expressed in Fourier space, the ILCF is defined by
\begin{align}
  \label{eq:ilcf-definition}
  l(r)\,\equiv\, &\frac{V^3}{(2\pi)^{3D}}\left(\frac{r^D}{V}\right)^{3/2} \hspace*{-1em}\underset{{|\B{k}_1|,|\B{k}_2|,|\B{k}_3| \leq
      2\pi/r}}{\int\hspace{-0.25em}\int\hspace{-0.25em}\int}\hspace{-1em} \D{^D\B{k}_1}\,\D{^D\B{k}_2}\,\D{^D\B{k}_3} \nonumber \\[0.5em]
  &\times\,\text{e}^{i[\B{k}_1 \cdot \B{x}+\B{k}_2 \cdot (\B{x}+\B{r})+\B{k}_3 \cdot (\B{x}-\B{r})]}\,\left<\epsilon_{\B{k}_1}\,\epsilon_{\B{k}_2}\,\epsilon_{\B{k}_3}\right>\,,
\end{align}
where $D$ denotes the dimension of space, $V$ the volume of the survey and $\left<\,\right>$ an ensemble average over all realisations of the density field. In writing
Eq.~(\ref{eq:ilcf-definition}) we adopted the same conventions as employed in \cite{Wolstenhulme:2014cla}.

The ILCF differs from the conventional three-point correlator of whitened density fields in two respects. Firstly, it is multiplied by a prefactor $(r^D/V)^{3/2}$ and
secondly, each field is convolved with a top-hat window function that cuts off high-frequency modes $|\B{k}| > 2\pi/r$. These corrections are introduced to regularize the
integral in the regimes of large and small scales, respectively. As discussed in \cite{linecorr}, Fourier modes which are associated with scales larger than the longest
physical correlation length or scales smaller than the lowest characteristic length obtain random phase factors that would cause the ILCF to be dominated by random noise.

It is important to reiterate that the ILCF is independent of amplitude information and for that reason not just a regularized integral of the bispectrum. It depends on the
quantity $\left<\epsilon_{\B{k}_1}\,\epsilon_{\B{k}_2}\,\epsilon_{\B{k}_3}\right> =
\left<\exp{\left[i\left(\theta_{\B{k}_1}+\theta_{\B{k}_2}+\theta_{\B{k}_3}\right)\right]}\right>$, which can be determined by using the probability density function (PDF) of
Fourier phases, ${\cal P}\{\theta\}$. This PDF was computed perturbatively for mildly non-Gaussian fields by \cite{Matsubara:2003te}, showing that it is not solely related to
the bispectrum but a progression of all higher order spectra such as the trispectrum etc. Taking this result, \cite{Wolstenhulme:2014cla} showed that the ILCF can be expressed
to lowest order as follows:
\begin{align}
  \label{eq:ilcf-perturbative}
  &l(r)\,\simeq\, \left(\frac{\sqrt{\pi}}{2}\right)^3\bigg(\frac{r}{2\pi}\bigg)^{3D/2} \hspace*{-1em}\underset{{|\B{k}_1|,|\B{k}_2|,|\B{k}_1+\B{k}_2| \leq
      2\pi/r}}{\int\hspace{-0.25em}\int}\hspace{-1em} \D{^D\B{k}_1}\,\D{^D\B{k}_2} \nonumber \\[0.5em]
  &\times\,\frac{B(\B{k}_1,\B{k}_2)}{\sqrt{P(|\B{k}_1|)\,P(|\B{k}_2|)\,P(|\B{k}_1+\B{k}_2|)}}\,\cos{\left[\left(\B{k}_1-\B{k}_2\right) \cdot
      \B{r}\right]}\,,  \\[-1em] \nonumber
\end{align}
where $P(|\B{k}|)$ and $B(\B{k}_1,\B{k}_2)$ are the ordinary power spectrum and bispectrum. Accordingly, we can think of the ILCF as compressing information from all higher
order spectra into a single function.  Only in the case of weakly non-Gaussian fields does the dominant contribution stem from the bispectrum.

For evaluating the ILCF on a given density field, Eq.~(\ref{eq:ilcf-perturbative}) is not particularly useful. However, we arrive at a simple prescription to extract the line
correlation by assuming ergodicity and replacing the ensemble average in Eq.~(\ref{eq:ilcf-definition}) by an average over all translations and rotations. Integrating out the
translation vector $\B{x}$ as well as the orientation of $\B{r}$ thus gives
\begin{align}
  \label{eq:ilcf-prescription}
  l(r)\,=\, &\frac{V^2}{(2\pi)^{2D}}\left(\frac{r^D}{V}\right)^{3/2} \hspace*{-1em}\underset{{|\B{k}_1|,|\B{k}_2|,|\B{k}_1+\B{k}_2| \leq
      2\pi/r}}{\int\hspace{-0.25em}\int}\hspace{-1em} \D{^D\B{k}_1}\,\D{^D\B{k}_2} \nonumber \\[0.5em]
  &\times\,w_D(|\B{k}_1-\B{k}_2|\,r)\,\frac{\delta_{\B{k}_1}\,\delta_{\B{k}_2}\,\delta_{-\B{k}_1-\B{k}_2}}{|\delta_{\B{k}_1}\,\delta_{\B{k}_2}\,\delta_{-\B{k}_1-\B{k}_2}|}\,,
\end{align}
where we have defined the rotational average of the exponential Fourier factor as
\begin{align}
  \label{eq:wd}
  \hspace*{-0.25em}w_D(k\,r) \equiv \left<\text{e}^{i\B{k}\cdot \B{r}}\right>_{\cal R} = \left\{\begin{array}{ll} \cos{(k\,r)}\,, & \text{if $D=1$ ,} \\ J_0(k\,r)\,, &
      \text{if $D=2$ ,} \\ \sin{(k\,r)}/(k\,r)\,, & \text{if $D=3$ .}\end{array}\right.
\end{align}
A discretized version of Eq.~(\ref{eq:ilcf-prescription}) will be used for all numerical computations throughout this paper. Details on the implementation are given in
App.~\ref{sec:app.implementation}.

\subsection{Properties and Applications}
\label{sec:ILCF.properties+applications}

The morphology of the cosmic web, composed of structures such as clusters, filaments and voids, is largely encoded in the phases of Fourier coefficients \citep[for a
demonstration see e.g. ][]{chiang2000}. Against this background, it is natural to assume that the ILCF is sensitive to these kinds of objects and, owing to its linear
configuration, especially to filaments. In fact, after a more systematic analysis, \cite{linecorr} conclude that the ILCF is generally a measure for aspherical structures on
scales $\sim 2r$. This holds true for oblate and prolate objects alike, so that $l(r)$ is able to probe both cosmic sheets and filaments.

Furthermore, they find two scaling relations depending on the size and density of the substructure. Letting $r_0$ denote its characteristic scale and $N$ the number of
objects\footnote{All Objects are assumed to have random positions and orientations.}, their filling factor is given by $f = N\,V_0/V$ where $V_0$ is the volume of a
$D$-dimensional sphere of radius $r_0$. A change of the characteristic scale or the number of objects has the following effect on the ILCF:
\begin{alignat}{2}
  &a) \hspace*{1em} \left.\begin{array}{ll} r_0 &\hspace{-0.75em}\rightarrow \alpha\,r_0 \\[0.2em] f &\hspace{-0.75em}= \text{const.}
    \end{array}\right\} \hspace{2em} &&\Rightarrow \hspace{2em} l(r) \rightarrow l(r/\alpha) \,,\\
  &b) \hspace*{2em} N \rightarrow \alpha\,N \hspace{2em} &&\Rightarrow \hspace{2em} l(r) \rightarrow \frac{l(r)}{\sqrt{\alpha}}\,.
\end{alignat}
The first relation specifically targets the behaviour of $l(r)$ under a rescaling of the substructure and therefore should not be confused with a scaling law resulting from
power law clustering. This can be derived from Eq.~\ref{eq:ilcf-perturbative} and assuming the hierarchical model \citep[e.g. ][]{Bernardeau2002}, so that we have $B \sim
P^2$. For a power law spectrum $P(k) \sim k^n$ we then get
\begin{align}
  l(\alpha\,r) = \alpha^{-\frac{D+n}{2}}\,l(r)\,,
\end{align}
implying that the slope of $l(r)$ is less steep than that of the two-point function, which would scale as $\alpha^{-(D+n)}$ for the same power spectrum. This conclusion was
already found from numerical observations in \cite{linecorr}. Furthermore, compared to conventional correlation functions, the second relation is also somewhat unintuitive:
increasing the number of objects (and hence the clustering) causes a decrease in the amplitude of $l(r)$ instead of an increase, as one would expect. The cause of this feature
is that the increased number of different translations and rotations of objects randomize the phase factors, hence decreasing their correlation. This effect can also be
understood by employing the halo model for a particular class of objects with orientation $\B{\phi}$, mass $m$ and density profile $\delta_H(\B{r}\,|\,\phi,m)$ with Fourier
transform $U_{\B{k}}(\B{\phi},m)$. For random positions of the halos, one only needs to consider the one-halo term, so that, according to \cite{Smith2005}, the power- and
bispectrum become
\begin{align}
  \label{eq:halomodel-powerspec}
  \frac{P(|\B{k}|)}{(2\pi)^D} &= \int \D{m}\,\frac{n(m)\,m^2}{\overline{\rho}^{\,2}} \int \frac{\D{\B{\phi}}}{\Omega_D}\,|U_{\B{k}}(\B{\phi},m)|^2\,, \\
  \label{eq:halomodel-bispec}
  \frac{B(\B{k}_1,\B{k}_2,\B{k}_3)}{(2\pi)^D} &= \int \D{m}\,\frac{n(m)\,m^3}{\overline{\rho}^{\,3}} \int \frac{\D{\B{\phi}}}{\Omega_D}\,\prod_{i=1}^3 U_{\B{k}_i}(\B{\phi},m)\,,
\end{align}
where $\overline{\rho}$ denotes the total mass density of all halos and $\Omega_D$ the surface area of the $D$-dimensional unit sphere. If all halos have identical mass $M$,
the mass function is given by $n(m) = N/V\,\delta^D(m-M)$ and we see that $P \propto 1/N$, i.e. $B \propto 1/N^2$. Hence, the combination of power spectra and the bispectrum
in Eq.~(\ref{eq:ilcf-perturbative}) gives rise to the scaling $\sim 1/\sqrt{N}$.

Being a relatively new tool in large-scale structure statistics, the ILCF has been applied only a few times so far. For instance, in \cite{linecorr} it was shown to be a
significantly more sensitive tool to analyse properties of dark matter than the two-point function. The authors examined a series of cosmological N-body simulations with
different warm dark matter masses, finding that $l(r)$ distinguishes those masses out to scales approximately five times larger than the two-point function. This feature is a
direct consequence of the ILCF's sensitivity to elongated structures and its ability to exclusively probe the non-linear stages of gravitational evolution. In another recent
application \citep{Alpaslan:2014ura}, the ILCF was used to identify a possibly new class of structure in the matter distribution, denoted as \emph{tendril galaxies}. These
galaxies were spotted in the Galaxy and Mass Assembly (GAMA, \cite{driver2009}) catalogue as remnants after all filament and isolated galaxies had been removed by appropriate
detection algorithms. They appear in thin chains that extend into voids and display a different line correlation than those galaxies residing in the larger filaments. Finally,
in \cite{Wolstenhulme:2014cla}, the ILCF was computed using tree-level perturbation theory giving good agreement with direct estimates from N-body simulations. An analytic
formula for $l(r)$ allows it to be modelled properly and opens the door for measuring various cosmological parameters.


\section{The Anisotropic Line Correlation Function}
\label{sec:ALCF}

Our ability to measure cosmological parameters with the line correlation function is intimately linked to our understanding of how it is influenced by effects that cause
anisotropies in the distribution of galaxies (see Sec.~\ref{sec:RSD}). Not knowing the systematics of these effects increases uncertainties in the result. However, an
anisotropic signal does not only add complications, its detection would also open up new avenues to probe and constrain the cosmological evolution in our universe. For this
reason, it is a vital step to extend the ILCF to account for radial and transverse distance scales separately, allowing us to detect distortions along the line-of-sight.

Due to the corrections regularising the ILCF, there are various possibilities to break rotational symmetry. In the following we will therefore explore two different mode
cut-offs, one being spherical as in the definition in Eq.~(\ref{eq:ilcf-definition}), the other one aspherical. We show that the latter is better suited to pick up
anisotropies and test some of its properties.

\begin{figure}
  \centering
  \resizebox{0.70\columnwidth}{!}{\input{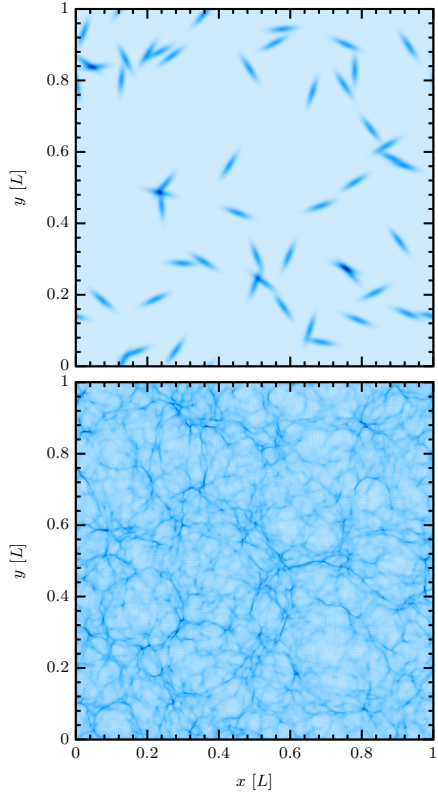}}
  \caption{Examples of two-dimensional density fields used throughout the paper. Upper panel: $50$ aspherical Gaussian halos with $\sigma_x = 0.006\,L$, $\sigma_y = 0.024\,L$ 
    and orientations drawn from a flat distribution. Lower panel: Zel'dovich approximation, cosmological parameters are to be found in the text.}
  \label{fig:fields}
\end{figure}

\subsection{A Spherical Cut-Off}
\label{sec:ALCF.spherical}

Naively, it is straightforward to write down a two-dimensional analogue of the ILCF, especially when adopting the distant-observer approximation, so that we can identify the
line-of-sight direction with the $z$-direction -- we simply restrict the angular average in our prescription in Eq.~(\ref{eq:ilcf-prescription}) to orientations of the
transverse part of the vector $\B{r}$, giving
\begin{align}
  \label{eq:alcf-sph-prescription}
  l(r_\perp,r_\parallel)\,&=\, \frac{V^2}{(2\pi)^{2D}}\left(\frac{r^D}{V}\right)^{3/2} \hspace*{-1em}\underset{{|\B{k}_1|,|\B{k}_2|,|\B{k}_1+\B{k}_2|
      \leq 2\pi/r}}{\int\hspace{-0.25em}\int}\hspace{-1em} \D{^D\B{k}_1}\,\D{^D\B{k}_2} \nonumber \\[0.5em]
  &\times\, \cos{\left[(k_{\parallel,1}-k_{\parallel,2})\,r_\parallel\right]}\,w_{D-1}(|\B{k}_{\perp,1}-\B{k}_{\perp,2}|\,r_\perp)\, \nonumber \\[0.5em]
  &\times\, \frac{\delta_{\B{k}_1}\,\delta_{\B{k}_2}\,\delta_{-\B{k}_1-\B{k}_2}}{|\delta_{\B{k}_1}\,\delta_{\B{k}_2}\,\delta_{-\B{k}_1-\B{k}_2}|}\,.
\end{align}
Here, $r_\perp \equiv \sqrt{x^2+y^2}$ and $r_\parallel \equiv z$ $(D=3)$ denote transverse and radial separations, respectively. 

\begin{figure}
  \centering
  \resizebox{0.98\columnwidth}{!}{\input{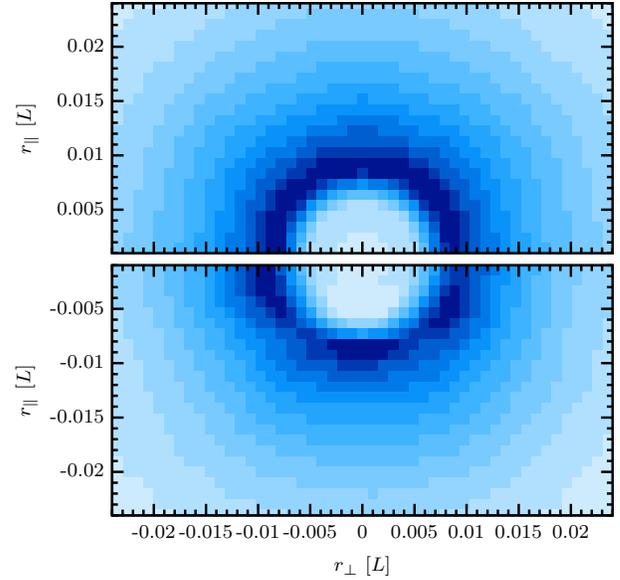}}
  \caption{Anisotropic line correlation function with spherical cut-off for statistically isotropic (top) and anisotropic fields with von Mises parameters $\kappa = 3$, $\mu =
    0$ (bottom). Each field contains $50$ halos and the width of their Gaussian profiles are chosen as  $\sigma_x = 0.006\,L$ and $\sigma_y = 0.024\,L$. Results are
    averaged over $50$ different realisations. Colour scales are adjusted.}
  \label{fig:SPHvM2d}
\end{figure}

To see how well the function above distinguishes between statistically isotropic and anisotropic density fields, we consider a simple test case. We construct fields consisting
of superpositions of aspherical halos with a Gaussian density profile (see upper panel in Fig.~\ref{fig:fields}). While their centre positions are randomly distributed, their
orientations are either drawn from a flat or a von Mises probability distribution. The von Mises distribution \citep{mardia} is a close approximation of the \emph{wrapped
  normal distribution} -- the analogue of the usual normal distribution on a circle -- and depends on two parameters: $\mu$, the centre of the distribution; and $\kappa$,
which is a measure for its `peakedness', i.e.~increasing $\kappa$ causes a higher concentration about the centre. It reads
\begin{align}
  \label{eq:vonMises}
  p(\phi\,|\,\mu,\kappa) = \frac{\text{e}^{\kappa\,\cos{\left(\phi-\mu\right)}}}{2\pi\,I_0(\kappa)}\,, \\[-1.25em] \nonumber
\end{align}
where $I_0(\kappa)$ denotes the modified Bessel function of zeroth order. Furthermore, all fields here and in the remainder of this paper, if not stated otherwise, are set up
in a two-dimensional box of sidelength $L$ with $N=512$ grid cells (details are given in the captions of the corresponding figures). Extracting the line correlations of these
sample fields according to Eq.~(\ref{eq:alcf-sph-prescription}) leads to the plot in Fig.~\ref{fig:SPHvM2d}. While $l(r_\perp,r_\parallel)$ only depends on $r =
\sqrt{r_\parallel^2+r_\perp^2}$ for the statistically isotropic fields, as expected, we do not observe a clear deviation from rotational symmetry in case of the anisotropic
ones. This implies that the anisotropic signal is smeared over the whole $(r_\perp,r_\parallel)$-plane, which renders the spherical cut-off either tricky or ill-suited for the
purposes given in the introduction of this section. For that reason we consider an alternative method that incorporates an aspherical cut-off.

\subsection{An Aspherical Cut-Off}
\label{sec:ALCF.aspherical}

\begin{figure}
  \centering
  \resizebox{\columnwidth}{!}{\input{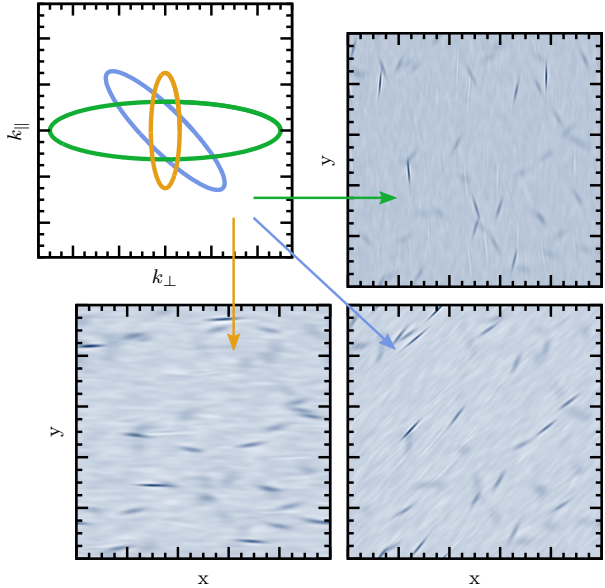}}
  \vspace*{-2em}
  \caption{Cut-off ellipses for different pairs $(r_{\perp},r_{\parallel})$ and the resulting whitened and filtered density fields. Note that the blurring increases from top
    right to bottom left because of a decreasing number of modes.}
  \label{fig:cutoff}
\end{figure}

When exploiting our freedom to alter the mode cut-off, we must still ensure that it regularises the integral for large scales and at the same time does not introduce any
artificial anisotropies. Therefore, we impose that it should satisfy the following two conditions:
\begin{enumerate}[labelindent=*,style=multiline,label=(\arabic*),leftmargin=*]
\item The number of enclosed modes scales with $r^{-D}$;
\item The number of enclosed modes for each pair of values $(r_\perp,r_\parallel)$ with the same $r$ is constant.
\end{enumerate}
There are still many possible cut-offs that satisfy these conditions; the one that we propose here is based on oblate spheroids of constant eccentricity (i.e.~ellipses in the
two-dimensional $(k_\perp,k_\parallel)$-plane). The semi-minor axes is supposed to scale as $2\pi/r$ and more importantly, its orientation is chosen to be aligned with the
vector $\B{r} = (\B{r}_\perp,r_\parallel)$, such that the spheroid rotates for varying scales. This choice is motivated by the fact that by filtering out all modes except
those that predominantly belong to a given direction in $k$-space, objects are singled out that are aligned with the transverse direction in real space. This is demonstrated
for a two-dimensional example in Fig.~\ref{fig:cutoff} which shows three different cut-offs and the corresponding whitened and filtered density fields. The green ellipse, for
instance, belongs to a configuration where $r_\perp = 0$.
Mathematically, this \emph{rotating} cut-off may be expressed as
\begin{align}
  \label{eq:ROT} \theta_\eta(\B{k}, \B{r}) \equiv k^2\,r^2 + \left(\eta^2-1\right) \left(\B{k} \cdot \B{r}\right)^2 \leq 4\,\pi^2\,,
\end{align} 
where $\eta = \text{const.} \geq 1$. In two dimensions, we can convert Eq. (\ref{eq:ROT}) into polar form with $\varphi$ the polar angle in the $(k_\perp,k_\parallel)$-plane
and we get
\begin{align} 
  k \leq \frac{2\,\pi/(\eta\,r)}{\sqrt{1-\left(1-\frac{1}{\eta^2}\right)\left(\frac{r_{\parallel}}{r}\cos{\varphi}-\frac{r_{\perp}}{r}\sin{\varphi}\right)^2}}\,.
\end{align}
Thus, both semi-axes scale with $r^{-1}$ while the eccentricity of the ellipses is constant and given by $\varepsilon = \sqrt{1-\eta^{-2}}$, as desired. For $\eta = 1$ we
recover the original spherical truncation, while different values of $\eta$ influence the signal-to-noise ratio and the sensitivity to anisotropies, as we will see in the
following.

\begin{figure}
  \centering
  \resizebox{\columnwidth}{!}{\input{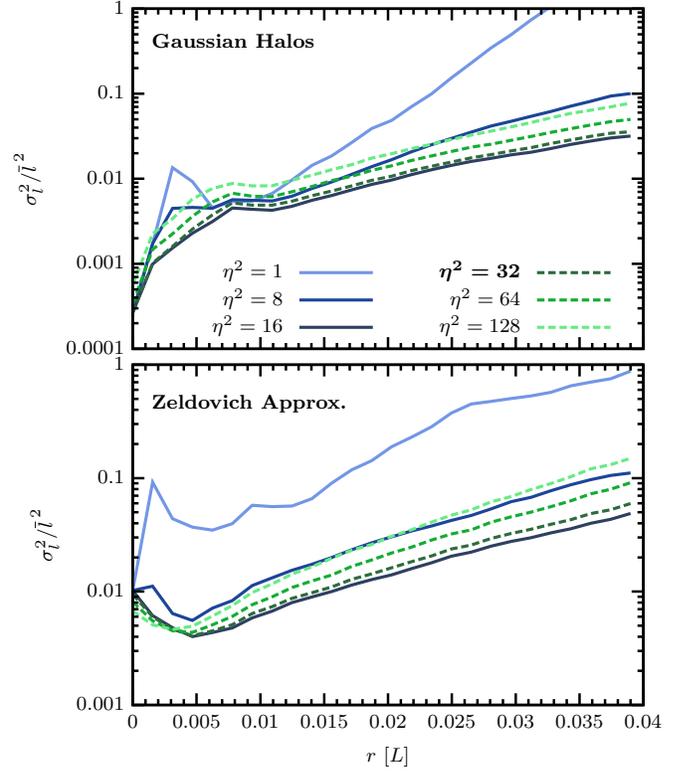}}
  \caption{Ratio of variance to squared average as a function of radial scale $r$ based on $20$ realisations. Upper panel: density fields consisting of Gaussian halos with
    random positions and orientations. Lower panel: fields set up according to the Zel'dovich approximation, also based on $20$ realisations.}
  \label{fig:eta}
\end{figure}

Since the eccentricity is left as a free parameter, we may ask whether there is a particular choice of $\eta$ that is optimal in the sense that it gives optimal
signal-to-noise ratios for a range of scales. To determine if there is such a choice we employ the same sample of statistically isotropic fields as above, consisting of
Gaussian halos with random positions and orientations. After computing the average line correlation and its variance from a number of realisations, we integrate out the
angular dependency by averaging over all data points belonging to one of $25$ radial bins of width $\Delta r \sim 0.0016\,L$ that extend outwards from the origin
$r=0$. Normalizing the variance by the square of the average, we obtain an $r$-dependent noise-to-signal measure that we can compare for different values of $\eta$. The
results of this analysis are shown in the upper panel of Fig.~\ref{fig:eta}. All curves display a nearly continuous increase in the noise level, which is a consequence of a
decreasing number of modes. However, the line correlation with spherical cut-off exhibits the largest noise-to-signal for most scales, indicating that a different choice for
$\eta$ is preferable. Indeed, we observe that $\sigma_l^2/\overline{l}^{\,2}$ drops when the cut-off becomes more elliptical and eventually reaches a minimum that roughly lies
in the interval $\eta^2 \sim [16,32]$. For larger values, the noise level begins to rise again, such that the optimal $\eta$ must reside in the given interval. However, since
the geometry of the cut-off matches that of the Gaussian halos, one might suspect that this choice of parameter is optimal for this particular class of density fields only. To
check this suspicion, we employ a second, more realistic set of density fields which are set up according to the Zel'dovich approximation \citep{zeldovich1970} for a
$\Lambda$CDM universe with parameters $\Omega_m = 0.314$, $h = 0.674$, $\sigma_8 = 0.9$ and baryon fraction $f_b = 0.038$ in a box of sidelength $L = 1\,h^{-1}\,$Gpc (for an
example see lower panel of Fig.~\ref{fig:fields}). Applying the same analysis to the Gaussian halo fields leads to the results in the lower panel of Fig.~\ref{fig:eta}. As
before, we note that a minimal noise-to-signal ratio is obtained in the parameter range $\sim [16,32]$, strengthening the previous result\footnote{This result was obtained in
  a $2D$ analysis; we caution the reader that the optimal value for $\eta$ in $3D$ might be different.}. Therefore, we define the anisotropic line correlation function (ALCF)
for the remainder of this paper by adopting the rotating cut-off with $\eta^2 = 32$. After fixing $\eta$, we still have the freedom to adjust the profile of the cut-off, for
instance to a Gaussian instead of a top-hat profile, and thus minimize the noise-to-signal ratio even further. However, for simplicity, we work with the top-hat filter in this
article.

\begin{figure}
  \centering
  \resizebox{\columnwidth}{!}{\input{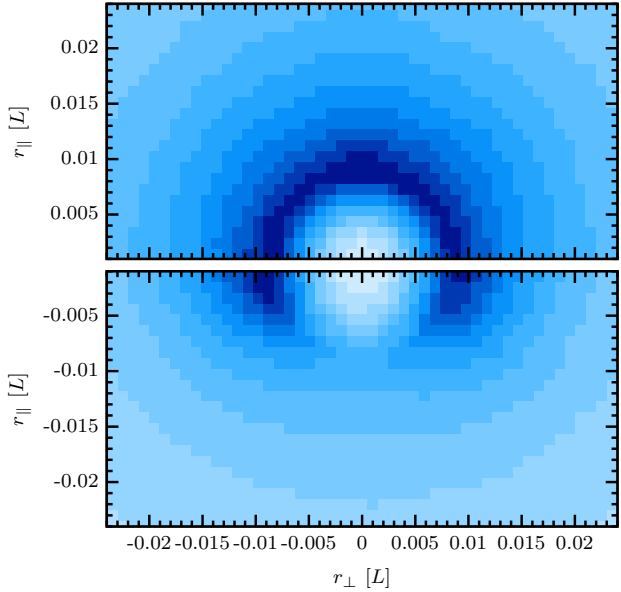}}
  \caption{Same as Fig.~\ref{fig:SPHvM2d} but using the anisotropic line correlation function with aspherical, rotating cut-off. Colour scales are adjusted.}
  \label{fig:ROTvM2d}
\end{figure}

Having settled on a definition of the ALCF, let us see how this choice influences the sensitivity to anisotropies. To that end, we repeat the test case from above where the
fields are either statistically isotropic or anisotropic by giving the orientation of the Gaussian halos a strong preference in the line-of-sight direction. The resulting
ALCFs are displayed in Fig.~\ref{fig:ROTvM2d}. While the isotropic fields still give rise to rotationally invariant line correlations, we now see a clear anisotropic signal in
the lower panel. Since the isocontour lines appear squashed and the signal peaks close to $r_\parallel = 0$, the whole transverse direction is enhanced. Given that most of the
halos point in the line-of-sight direction, this feature may appear counter-intuitive, but it is simply a continuation of the properties of the ILCF mentioned earlier. In
Sec.~\ref{sec:ILCF.properties+applications} it was found that the amplitude of the ILCF decreases when the number of objects increases, which is reproduced here as well. In
the following section we study the properties of the ALCF more closely.

\subsection{Properties and Examples}
\label{sec:ALCF.properties}

To characterise the properties of the ALCF and verify that it extends those of the ILCF in a consistent manner, we work again with the simple Gaussian halo fields and consider
two anisotropic setups. For the first, we introduce a preferred direction, as already seen in the examples of the last two sections; for the second, we vary the scale of the
halos depending on their orientation.

\begin{figure}
  \centering
  \resizebox{\columnwidth}{!}{\input{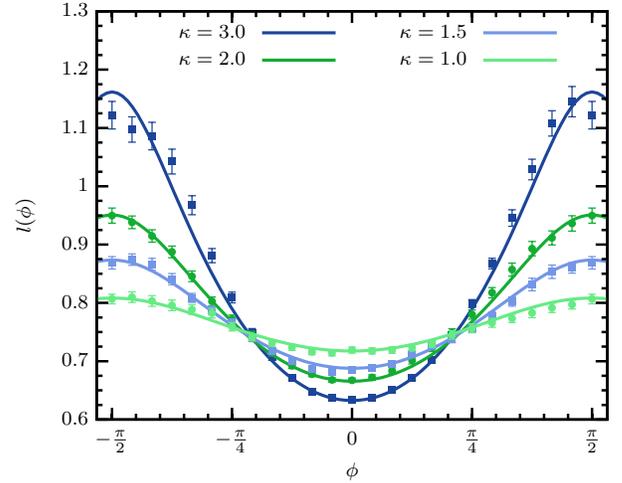}}
  \caption{Angular dependence of the ALCF for von Mises dsitributed density fields. Data points stem from the extraction procedure described in the text with error bars
    representing the $1$-$\sigma$ uncertainty from averaging over $100$ realisations. Solid lines are the corresponding fitting results.}
  \label{fig:vMfit}
\end{figure}
\begin{table}
  \centering
  \caption{Input and fit parameters for von Mises distributed density fields. Parameters $\alpha$ and $\beta$ are fitted simultaneously to all data sets, giving 
    $\alpha=0.58\pm 0.02$ and $\beta=-0.23\pm 0.02$.}
  \label{tab:vmFit}
  \vspace*{0.25em}
  \setlength{\tabcolsep}{0.5em}
  \begin{tabular}[!b]{lccccc}
    \hline \hline \\[-1em]
    & $\kappa_1$ & $\kappa_2$ & $\kappa_3$ & $\kappa_4$ & $\kappa_5$ \\\\[-1em] \hline \\[-0.5em]
    Input & $0.5$ & $1.0$ & $1.5$ & $2.0$ & $3.0$ \\
    Fit   & $0.5\pm 0.1$ & $1.1\pm 0.1$ & $1.7\pm 0.1$ & $2.3\pm 0.2$ & $3.4\pm 0.2$  \\\\[-0.5em] \hline    
  \end{tabular}
  \vspace*{0.1em}
\end{table}
\begin{figure*}
  \centering
  \resizebox{1.89\columnwidth}{!}{\input{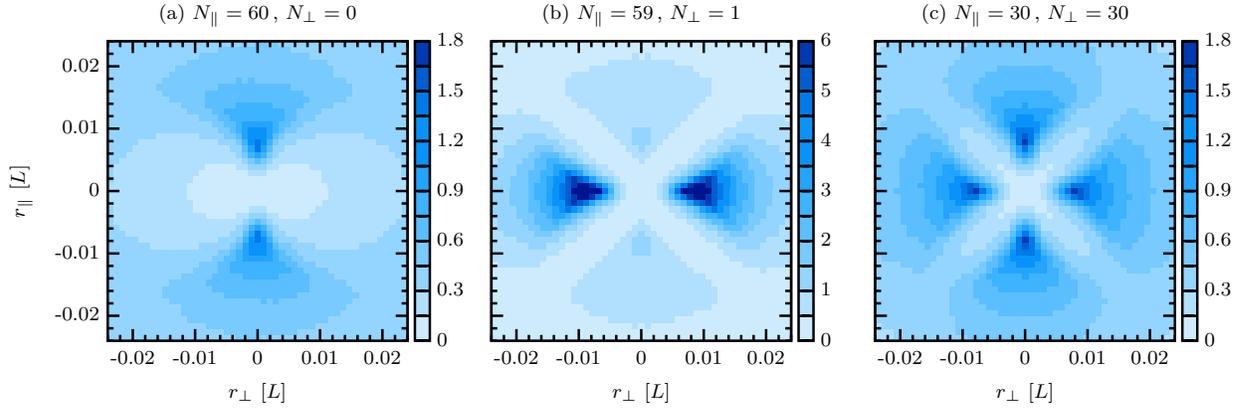}}
  \caption{Line correlations for fields with different numbers of elongated Gaussians pointing either along the line-of-sight ($N_\parallel$) or transverse direction ($N_\perp$),
    based on $20$ realisations each. Note the change of colour scale in the middle panel.}
  \label{fig:rtGaussians}
\end{figure*}

\subsubsection{Varying Number Densities} 
\label{sec:ALCF.properties.numberdensities} 

As before, the preferred direction is established by drawing the orientation angles of halos from a von Mises distribution. We want to analyse how the strength of anisotropy,
parametrised by the von Mises parameter $\kappa$, affects $l(r_\perp,r_\parallel)$ and in particular, whether there is a scaling relation similar to the one valid for the
ILCF. Therefore, we firstly introduce a probability density of objects, $p_H(\phi)$, which is defined on the interval $[-\pi/2,\pi/2]$, such that \begin{align} N\times
  p_H(\phi)\,\D{\phi} \equiv \text{number of objects in the bin } \D{\phi}\,,
\end{align}
\begin{align}
  \label{eq:normNumberDensity}
  \Rightarrow \hspace{2em} \int_{-\pi/2}^{\pi/2}p_H(\phi)\,\D{\phi} = 1 \\[-1.25em] \nonumber
\end{align}
with $N$, the total number of objects. Due to the symmetric geometry of the halos, $p_H$ is related to the probability density of orientation angles by
\begin{align}
  p_H(\phi) = p(\phi)+p(\phi+\pi) = \frac{\cosh{\left[\kappa\,\cos{(\phi-\mu)}\right]}}{\pi\,I_0(\kappa)}\,,
\end{align}
where we have substituted in the von Mises distribution from Eq.~(\ref{eq:vonMises}). 

As already suggested by Fig.~\ref{fig:ROTvM2d} and in agreement with the interpretation of line correlations, we expect that the ALCF scales inversely with the probability
density of objects. To verify this assertion, we compute the ALCF for five ensembles, each with $\mu=0$ and a different value for $\kappa$. Afterwards, we extract the
dependence on the angle $\phi$ from the two-dimensional function $l(r_\perp,r_\parallel)$ by averaging over a fixed number of points with constant $\phi$, but constrained to
the radial interval $0 \leq r \leq R=0.02\,L$ to exclude areas of large sample variance, i.e.
\begin{align}
  l(\phi) \equiv \frac{1}{R} \int_0^R \D{r}\,l(r\sin{\phi},r\cos{\phi})\,.
\end{align}
The results are displayed in Fig.~\ref{fig:vMfit}; error bars originate from averaging over $100$ realisations. We then fit this data to determine the scaling behaviour and to
see whether we can recover the respective values of $\kappa$ that were used as input parameters for the construction of the fields with the model
\begin{align}
  l(\phi\,|\,\kappa, \alpha, \beta) = \alpha\,[\;p_H(\phi\,|\,\kappa)\,]^\beta\,,
\end{align}
where we allow for a varying amplitude and exponent. Since $\alpha$ and $\beta$ should be universal (independent of $\kappa$), we fit them simultaneously to all five data sets
while we leave $\kappa$ free to vary amongst them. The outcome of this fit along with the input parameters is summarised in Tab.~\ref{tab:vmFit}, and shown for four example
data sets as the thick solid lines in Fig.~\ref{fig:vMfit}. We note that the fitted curves reproduce well the behaviour of the extracted data and we recover all peakedness
parameters nearly within the error bounds. Moreover, taking the fitted values of $\alpha$ and $\beta$ to make a prediction for fields with a flat probability density
(i.e.~$p_H = 1/\pi$), we compute $l = 0.76\pm 0.03$ while getting $l = 0.753\pm 0.001$ numerically. Hence, we can confirm that the ALCF scales with the probability
distribution of objects where it is likely that the parameter $\beta$ depends on the eccentricity of the cut-off ellipses.

Before moving on to the second test case we briefly describe another interesting feature related to the probability density of structures. To that end, we generate fields that
can be considered as an extreme case of the von Mises fields above: varying numbers of elongated Gaussians pointing either in radial or transverse direction with fixed $N =
N_\perp + N_\parallel$. From Fig.~\ref{fig:rtGaussians} we observe that a strong signal in the line-of-sight is received when all objects are pointing in that direction, see
panel (a). Even so, this signal can be altered significantly if a single object is aligned with the transverse direction, see panel (b). Adjusting $N_\perp$ and $N_\parallel$
until equality leads to a symmetric signal, see panel (c). This feature of the ALCF can be explained along the same lines as before: in (a), no information on structures in
the transverse direction is present, while in (b) one object is present, which henceforth dominates due to minimal phase noise. Thus, care must be taken when using the plots
of the ALCF to eyeball the strength of anisotropy.

\subsubsection{Angularly Varying Scales}
\label{sec:ALCF.properties.angularScales}

\begin{figure}
  \centering
  \vspace*{0.50em}
  \resizebox{\columnwidth}{!}{\input{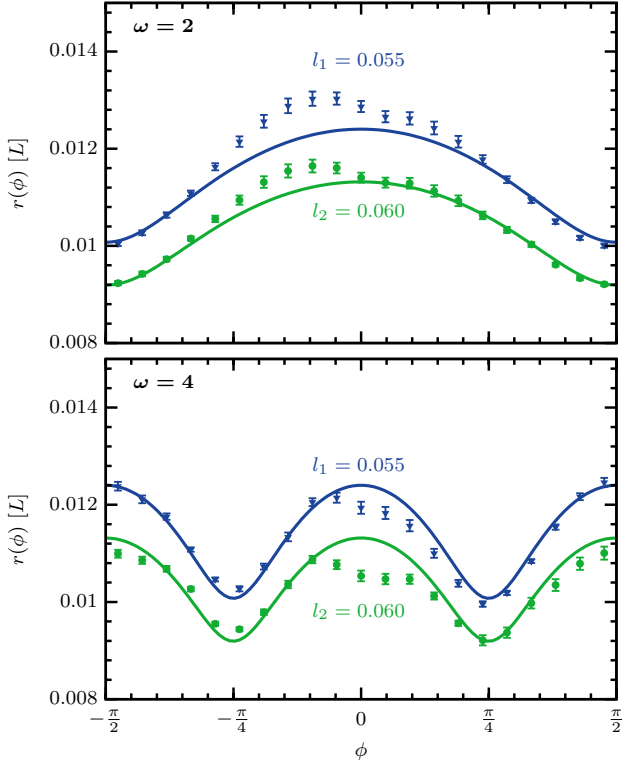}}
  \caption{Isocontours with $1$-$\sigma$ error bars for two different values of $l$ extracted from the ALCF of fields with angularly varying scales, based on $100$
    realisations each. Solid lines stem from the results of the fitting procedure.}
  \label{fig:vSfit}
\end{figure}

After having demonstrated the dependence on angular number densities, we turn our attention to angularly varying scales. Our sample halo fields are created as follows: for
each orientation angle $\phi \in [-\pi/2,\pi/2]$, we determine the scale $\sigma_y$ of the Gaussian profile from the function
\begin{align}
  {\cal S}(\phi\,|\,\overline{\sigma},\Delta \sigma, n) &= \overline{\sigma} + \Delta \sigma\,\cos{\left(\omega\,\phi\right)}
\end{align}
while keeping $\sigma_x = 0.006\,L$ fixed. As before, we intend to check how the ALCF is influenced by this anisotropy and especially how originally spherical lines of
constant $l$ are deformed. In order to suppress any additional variations of the amplitude of $l(r_\perp,r_\parallel)$, we need to hold the filling factor constant, so that
\begin{align}
  f(\phi) = p_H(\phi)\,\frac{V_0(\phi)}{L^D} \overset{\underset{\downarrow}{\hspace*{-0.2em}D = 2}}{=} \frac{\pi\,\sigma_x}{L^2}\,p_H(\phi)\,{\cal S}(\phi) =
  \text{const.}
\end{align}
\begin{align}
  \Rightarrow\hspace{1em} p_H(\phi) \propto 1/{\cal S}(\phi)\,,
\end{align}
where we resorted to the definition of $f$ in Sec.~\ref{sec:ILCF.properties+applications} and used $r_0 = \sqrt{\sigma_x\,\sigma_y}$ for $D=2$. Consequently, when determining
the scales from ${\cal S}$, the respective orientation angles must be drawn from the distribution $1/{\cal S}(\phi)$. This distribution is realised using the method of
\emph{inverse transform sampling} \citep{devroye}. With that, we set up two ensembles of fields for the cases $\omega=2$ and $\omega=4$, where we chose
$\overline{\sigma}=0.02\,L$ and $\Delta \sigma=0.01\,L$. After evaluating the line correlation function we extract two isocontours for each $\omega$, convert the Cartesian
coordinates into $21$ angular bins and average over $100$ realisations which gives the data points in Fig.~\ref{fig:vSfit}.

\begin{table}
  \centering
  \vspace*{0.25em}
  \caption{Expected and fitted parameters for density fields with scales of explicit angular dependence.}
  \setlength{\tabcolsep}{0.5em}
  \begin{tabular}[!h]{lcccc}
    \hline \hline \\[-1em]
    & $R_{l_1}$ [$10^{-2}\,L$] & $R_{l_2}$ [$10^{-2}\,L$] & $\Delta$ & $\gamma$ \\\\[-1em] \hline \\[-0.5em]
    Input & $1.14\pm 0.02$ & $1.03\pm 0.02$ & $0.5$ & $-$ \\
    Fit   & $1.16\pm 0.04$ & $1.06\pm 0.04$ & $0.65\pm 0.06$ & $0.13\pm 0.02$  \\\\[-0.5em] \hline    
  \end{tabular}
  \vspace*{0.1em}
  \label{tab:vsFit}
\end{table}

Since the average characteristic scale $\overline{r}_0 = \sqrt{\sigma_x\,\overline{\sigma}}$ is modified by the angle-dependent factor
$\sqrt{1+\Delta\,\cos{\left(\omega\,\phi\right)}}$ with $\Delta \equiv \Delta \sigma/\overline{\sigma}$, we may expect that any isocontour at scale $R$ for a statistically
isotropic field is distorted as
\begin{align}
  R \rightarrow R\,\left[1+\Delta\,\cos{\left(\omega\,\phi\right)}\right]^\gamma\,,
\end{align}
with an unknown exponent $\gamma$ which, as for the parameter $\beta$ in the previous section, presumably depends on the eccentricity of the cut-off ellipses. We therefore fit
the extracted contours with the model
\begin{align}
  r(\phi\,|\,R,\Delta,\gamma) = R\,\left[1+\Delta\,\cos{(\omega\,\phi)}\right]^\gamma
\end{align}
and leave $R$ free to vary while demanding that the parameters $\Delta$ and $\gamma$ shall be equal for all curves. The results of fitting them simultaneously to the data are
represented by the solid lines in Fig.~\ref{fig:vSfit}, with the corresponding best fit values given in Tab.~\ref{tab:vsFit}.

We observe that the data is fitted well by the model, showing the expected distortion of radial scales, i.e.~a stretching for those angles where the charactersitic scale is
larger. Furthermore, the fitted values for $R$ match well to the same isocontours of statistically isotropic fields, which were estimated from an ensemble of randomly
orientated halos with fixed scales $\sigma_x = 0.006\,L$ and $\sigma_y = \overline{\sigma} = 0.02\,L$. However, the parameter $\Delta$ is slightly off, which can be attributed
to the scatter in the data and the fact that the fit is quite sensitive on both $\Delta$ and $\gamma$.



\section{Application to Redshift-Space Distortions}
\label{sec:RSD}

Anisotropies in the distribution of galaxies occur naturally whenever the redshifts and positions (angles) acquired in a survey are converted into physical distances. They may
be of geometric or kinematic nature, denoted by the Alcock-Paczy\'{n}ski effect and (dynamical) redshift-space distortions, respectively. In this section, we analyse whether
these observational effects are picked up by the ALCF and how these results compare to the ones from conventional two-point statistics. Since we primarily aim for a proof of
concept, we will continue to use the two-dimensional halo and Zel'dovich mock fields introduced above.

\subsection{The Alcock-Paczy\'{n}ski Effect}
\label{sec:RSD.alcock-paczynski}

The conversion of the redshift difference and angular separation of a pair of galaxies to a physical distance depends on two basic cosmological quantities: knowledge of the
Hubble rate $H(z)$ is required for the computation of the line-of-sight separation; and the angular diameter distance $D_A(z)$ is required for the transverse separation,
\begin{align}
  \label{eq:APrp}
  r_\parallel &= \frac{\D{r}}{\D{z}}\,\Delta z = \frac{\D{}}{\D{z}}\,\left(\int_0^z \frac{c\,\D{z'}}{H(z')}\right)\,\Delta z = \frac{c\,\Delta z}{H(z)}\,,\\[0.5em]
  \label{eq:APrt}
  r_\perp &= D_A(z)\,(1+z)\,\Delta \theta\,.
\end{align}
Consequently, creating a spatial map of the universe hinges on a cosmological model that has to be assumed a priori; if this model differs from the true cosmology, the deduced
distances are wrong, causing distortions in the clustering signal. The first who proposed to use this effect as a probe for cosmological parameters were Alcock and
Paczy\'{n}ski \citep{alcock+paczynski} (AP). By considering an object whose intrinsic shape is known to be spherical, they deduced that measuring distortions from sphericity
can be used to constrain the combination $H(z)\,D_A(z)$, and in turn, for instance, the equation of state parameter of dark energy. However, in practice, there are no objects,
which are sufficiently spherical or whose length scales are known to sufficient accuracy, to perform the AP test on individually. Instead, one resorts to \emph{statistical
  standard rulers}, that is, length scales that are statistically imprinted on large-scale structures, such as the one set by baryon accoustic oscillations (BAO); since BAO's
are primarily a linear phenomenon, their signal is not considerably affected by non-linear physics. Thus, by measuring the position of the BAO peak and comparing it to the
theoretical prediction, it is possible to conduct an AP test \citep{Blake2003}. However, with the ever increasing volume coverage of galaxy surveys, an alternative has become
feasible, in which a complete model of a given statistical measure, incorporating a possibly wrong choice of cosmology, is fitted to the data (see
e.g. \citealt{Ballinger1996}, \citealt{okumura2008} or \citealt{padmanabhan2008}). A similar method could be employed based on the ALCF.

To quantify the impact of the AP effect on the line correlation function, we firstly parametrise the mismatch between true and assumed cosmology by two squashing factors
(following the notation of \citealt{Ballinger1996}):
\begin{align}
  \label{eq:APparameters}
  f_\parallel = \frac{H_a(z)}{H(z)}\,, \hspace{2em} f_\perp = \frac{D_{A}(z)}{D_{A,a}(z)}\,,
\end{align}
where here and in the following the subscript `$a$' indicates quantities derived from the assumed cosmology. From Eqs.~(\ref{eq:APrp}) and (\ref{eq:APrt}) we see that assumed
and true distance scales are related by the matrix
\begin{align}
  \label{eq:APscaling}
  \B{r}_a = {\cal S}\,\B{r}\,, \hspace{2em} {\cal S} = \left(
    \begin{array}[!h]{ccc}
      f_\perp^{-1} & & \\
      & f_\perp^{-1} & \\
      & & f_\parallel^{-1}
    \end{array}\right)\,.
\end{align}
Due to number conservation, the overdensity does not change in the assumed coordinate system, so that $\delta_a(\B{x}_a) = \delta(\B{x})$, which implies for the two-point
function
\begin{align}
  \label{eq:APxi}
  \xi_a(\B{r}_a) = \xi(\B{r}) = \xi({\cal S}^{-1}\,\B{r}_a)\,.
\end{align}
Thus, any squashing of galaxy separations directly translates to an equivalent squashing of the isocontours of $\xi$. To compute the analogue effect for the ALCF we need to be
more careful because of the scale dependent cut-off and prefactor. For that reason we transform $l_a(r_{\perp,a},r_{\parallel,a})$ piecewise and start from the whitened and
filtered density field in the assumed coordinate system, which we write as a convolution between $\epsilon$ and the cut-off $\Theta$ in real space,
\begin{align}
  \label{eq:aplcf1}
  \epsilon_a \ast \Theta_{a,\B{r}_a}(\B{x}_a) &= V_a\hspace{-1em}\int\displaylimits_{\theta(\B{k}_a,\B{r}_a) \leq 4\pi^2}\hspace{-1em} \D{^Dk_a}\,\text{e}^{i \B{k}_a \cdot
    \B{x}_a}\epsilon_{a,\B{k}_a} \nonumber \\
  &= V_a\,|{\cal S}|^{-1} \hspace{-2em} \int\displaylimits_{\theta({\cal S}^{-1}\B{k},\B{r}_a) \leq 4\pi^2}\hspace{-2em} \D{^Dk}\,\text{e}^{i \B{k} \cdot {\cal S}^{-1}\B{x}_a}\,\epsilon_{\B{k}}\,.
\end{align}
Here, we used that the Fourier modes scale inversely to spatial scales and employed the identity $\epsilon_{a,\B{k}_a} = \epsilon_{\B{k}}$, which is another direct consequence
of number conservation. Shifting the ${\cal S}$-matrix from $\B{k}$ to $\B{r}_a$ in the cut-off function causes an additional term that can be computed from the definition of
$\theta$ in Eq.~(\ref{eq:ROT}), yielding \vspace*{0.5em} \begin{align} \theta({\cal S}^{-1}\B{k},\B{r}_a) &= \theta(\B{k},{\cal S}^{-1}\B{r}_a) + \left(f_{\perp}^2 -
    f_{\parallel}^2\right)\,\left(k_{\perp}^2\,r_{a,\parallel}^2 - k_{\parallel}^2\,r_{a,\perp}^2\right) \nonumber \\ \label{eq:shift}
  &= \theta(\B{k},{\cal S}^{-1}\B{r}_a) + \delta_F(\B{k},{\cal S}^{-1}\B{r}_a)\,, \\[0.4em]
  \delta_F(\B{k},\B{r}) &\equiv (1-F^2)\,\left(F^{-2}k_{\perp}^2\,r_{\parallel}^2 - k_{\parallel}^2\,r_{\perp}^2\right)\,,
\end{align}
where $F$ is the ratio of $f_\parallel$ and $f_\perp$.\footnote{Note that $\delta_F$ arises purely as a consequence of the first term in Eq.~\ref{eq:ROT}. Eq.~\ref{eq:shift}
  and all of the following expressions are hence equally valid for the original spherical cut-off.} Plugging this result back into Eq.~(\ref{eq:aplcf1}) gives
\begin{align}
  \nonumber \epsilon_a \ast \Theta_{a,\B{r}_a}(\B{x}_a) &= V \hspace{-4.5em}\int\displaylimits_{\theta(\B{k},{\cal S}^{-1}\B{r}_a)+\delta_F(\B{k},{\cal S}^{-1}\B{r}_a) \leq 4\pi^2}\hspace{-4.5em}
  \D{^Dk}\,\text{e}^{i \B{k} \cdot {\cal S}^{-1}\B{x}_a}\,\epsilon_{\B{k}} \nonumber \\
  &\equiv \epsilon \ast \widetilde{\Theta}_{F,{\cal S}^{-1}\B{r}_a}({\cal S}^{-1}\B{x}_a)\,,
\end{align}
where we identified $V_a/|{\cal S}|$ with the true volume and let $\widetilde{\Theta}_F$ denote the altered cut-off that reverts to the original one for $F=1$. Having thus determined the relation between the whitened and filtered density field in the assumed cosmology to the one in the true model, we still need to transform the prefactor. Using Eq.~(\ref{eq:APscaling}) again, we arrive at \vspace*{0.5em} \begin{align}
  r_{a,\parallel}^2 + r_{a,\perp}^2 &= f_\parallel^{-2}\,r_\parallel^2 + f_\perp^{-2}\,r_\perp^2 =
  \left(f_\parallel^{-2}+f_\perp^{-2}\right)\,r_\parallel^2 + f_\perp^{-2}\,r^2 \nonumber  \\
  &= f_\perp^{-2}\,r^2\,\left[1+\left(F^{-2}-1\right)\,\mu^2\right]\,,
\end{align}
where $\mu$ is the cosine of the angle to the line-of-sight, i.e.~$\mu = r_\parallel/r$. All in all, we get
\begin{align}
  \label{eq:APALCF}
  l_a(\B{r}_a) = F^{\frac{3}{2}} \left[1+(F^{-2}-1)\left({\cal S}^{-1}\mu_a\right)^2\right]^{\frac{3D}{4}}\,\widetilde{l}_F\left({\cal S}^{-1}\B{r}_a\right)\,,
\end{align}
where
\begin{align}
  \left({\cal S}^{-1}\mu_a\right)^2 \equiv \frac{f_{\parallel}^2\,r_{a,\parallel}^2}{f_{\parallel}^2\,r_{a,\parallel}^2 + f_{\perp}^2\,r_{a,\perp}^2}\,.
\end{align}
Eq.~(\ref{eq:APALCF}), which is the main result of this section, relates the correlator for the assumed cosmology at scale $\B{r}_a$ to the modified (but true) one at the
shifted position ${\cal S}^{-1}\,\B{r}_a$. As in Eq.~(\ref{eq:APxi}) for the two-point function, it encodes the impact of the AP effect on the ALCF and can either be used as a
basis for fitting existing data to extract $f_\perp$ and $f_\parallel$ or for simulating it. Since the effect is not immediately evident by inspecting Eq.~(\ref{eq:APALCF}),
we compute the assumed line correlation function from a set of true fields with varying values for the squashing parameters.

To visualise the AP effect and enable comparison with the two-point function we decompose the data in the $(r_\perp,r_\parallel)$-plane into multipoles. As our density fields
are two-dimensional, the $n$-th order multipole is calculated via
\begin{figure}
  \centering
  \resizebox{\columnwidth}{!}{\input{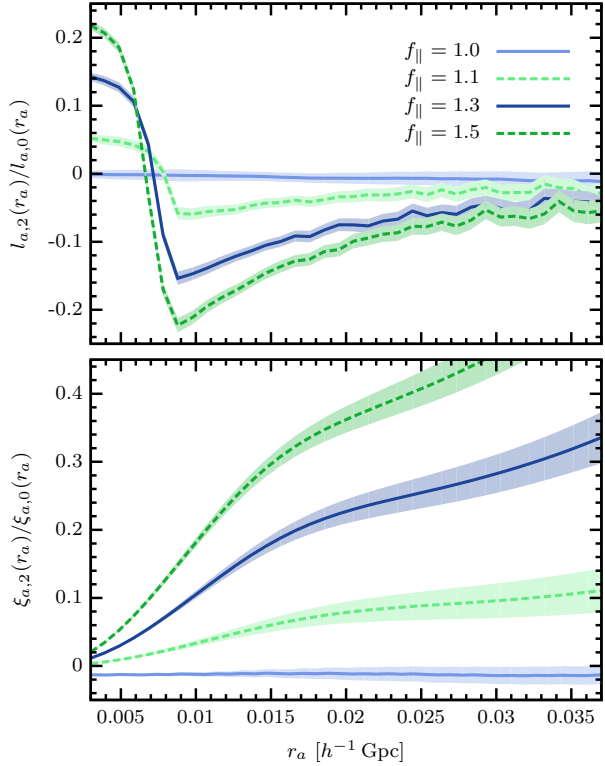}}
  \caption{Quadrupole-to-monopole ratio for the ALCF (upper panel) and two-point function (lower panel), derived from a set of $50$ Gaussian halo fields. The AP effect is
    simulated via Eq.~(\ref{eq:APALCF}) with $f_\parallel$ ranging between $1$ and $1.5$ while $f_\perp = 1$. Shaded areas mark $1$-$\sigma$ uncertainties.} 
  \label{fig:APGaussians}
\end{figure}
\begin{figure}
  \centering
  \resizebox{\columnwidth}{!}{\input{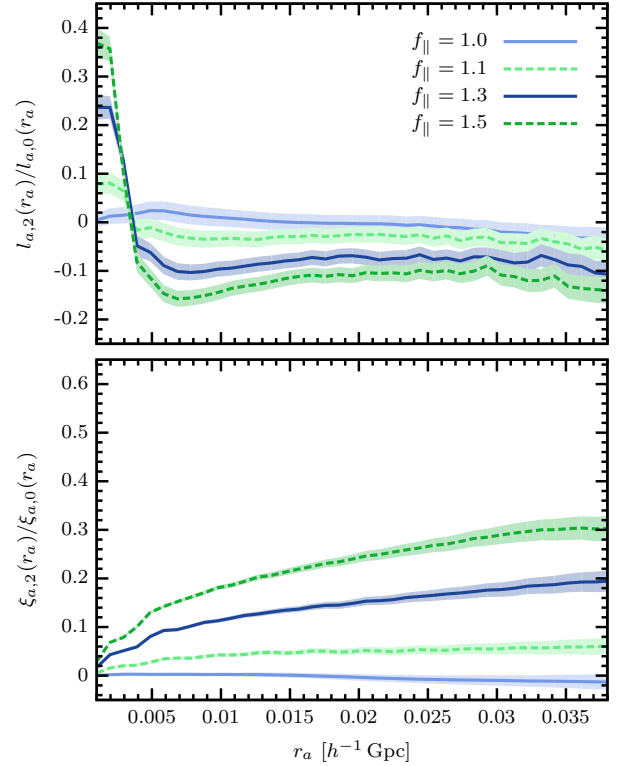}}
  \caption{Same as Fig.~\ref{fig:APGaussians}, but based on $50$ realisations of Zel'dovich density fields.}
  \label{fig:APZeldovich}
\end{figure}
\begin{align}
  \label{eq:multipole}
  f_n(r) = \frac{1}{\pi}\,\int_{-\pi}^{\pi}\,f(r\,\sin{\phi},r\,\cos{\phi})\,\cos{\left(n\,\phi\right)}\,\D{\phi}\,,
\end{align}
where $f$ is either $\xi$ or $l$ and the sign convention is chosen such that an enhancement of the transverse direction compared to the line-of-sight translates into a
positive signal for the quadrupole $f_2$. Using Eq.~(\ref{eq:APALCF}), we determine the monopole and quadrupole for an ensemble of Gaussian halo and Zel'dovich density fields
with varying $f_\parallel$, but fixed $f_\perp$. The corresponding quadrupole-to-monopole ratios are plotted in the upper panels of Fig.~\ref{fig:APGaussians} and
\ref{fig:APZeldovich}. We notice that in both cases the AP effect introduces a clearly visible anisotropy by means of an increasing quadrupole signal with rising
$f_\parallel$. Interestingly, the quadrupole is positive for small scales, meaning that the ALCF is squashed. However, with increasing scale the quadrupole changes sign,
turning the initial squashing into a stretching of the line-of-sight. After attaining a maximum, the amplitude of the quadrupole decreases, but remains negative throughout. In
comparison, the quadrupole-to-monopole ratio of the two-point function, which is displayed in the lower panels of Fig.~\ref{fig:APGaussians} and \ref{fig:APZeldovich}, is
positive for all scales, reflecting the squashing expected from Eq.~(\ref{eq:APxi}). Although being unintuitive, the behaviour of the ALCF can be understood in light of the
discussion in Sec.~\ref{sec:ALCF.properties}. If the separations of galaxies are squashed in a given direction, the scale and orientation of the structures in the density
field change. In the case considered here (squashing in the line-of-sight direction) a preference for structures to be aligned with the transverse direction arises. This
results in the enhanced radial signal since the ALCF scales inversely with the number density of objects.

While the ALCF is indeed capable of detecting the AP effect, the two-point function appears to be superior in telling apart different $f_\parallel$ from each other for both
sets of fields. Its quadrupole-to-monopole ratio increases with scale, which renders $\xi$ sensitive to $f_\parallel$ on all scales. On the other hand, the $1$-$\sigma$ error
regions begin to overlap at $r \sim 0.02\,h^{-1}\,$Gpc for the line correlation function, which implies that estimates based on $\xi$ will be statistically more significant.
This is a reasonable result owing to the number of modifications in Eq.~(\ref{eq:APALCF}), which tend to influence the ALCF in converse directions. Assuming $F > 1$ in the
following, we first note that the modified cut-off gives rise to an increase in the ALCF's amplitude in the line-of-sight direction, as can be verified
empirically. Conversely, the distortion of the prefactor attains a minimum for $r_{a,\perp} = 0$ and accordingly strengthens the transverse direction compared to the
line-of-sight. The same is true for the modification of scales by the matrix ${\cal S}^{-1}$, given that $l(\B{r})$ is a decreasing function of $|\B{r}|$. As a consequence,
the net effect of the AP squashing on the ALCF is smaller than for the two-point function, where only one kind of distortion effect is present.

However, we would like to reiterate that the ALCF uses phase information, which is complementary to the amplitude information retained in the two-point correlator. Thus, both
estimators can be used in conjunction in realistic applications to tighten constraints.

\subsection{Kinematical Redshift-Space Distortions}
\label{sec:RSD.kinematic}

Another consequence of measuring redshifts instead of physical distances are distortions due to the peculiar velocity of galaxies. Apart from the Hubble flow, they also
contribute to the redshift and thus cause the galaxies to appear displaced along the line-of-sight.

Since galaxies result from the structure formation process through gravity, their peculiar velocities are induced by gravity as well. On large scales, where structures have
not yet fully collapsed, all matter tends to fall into the nearest overdensity. For that reason, a given galaxy that resides on the near-edge of a large cluster tends to move
away from us, increasing its redshift. Consequently, its apparent position is closer to the center of the overdensity. A galaxy on the far-edge of the cluster would behave in
the opposite way, so that it appears to be closer to us than it actually qis. We therefore observe that structures on those large scales look squashed in radial direction
which is known as the \emph{Kaiser effect} \citep{1987MNRAS.227....1K}. He showed that in the distant observer limit, the power of scales that are aligned with the
line-of-sight is boosted by the factor $1+f\mu^2$ ($\mu$ being the cosine to the line of sight), so that the redshift-space dark matter power spectrum is given
by 
\begin{align}
  P_s(k) = (1+f\,\mu^2)^2\,P(k)\,.
\end{align}
The amplitude of the distortion depends on the logarithmic growth rate $f = \D{\ln{D}}/\D{\ln{a}}$, which is related to the matter content in the universe and, according to
\cite{peebles}, well approximated by the power law $f \approx \Omega_m^{0.6}$. Thus, measuring the Kaiser effect enables us to put constraints on the matter density, or, if
combined with an independent measurement of $\Omega_m$, on the theory of general relativity.

At much smaller, non-linear scales, objects within already collapsed structures acquire random virialized velocities which smear the structure in the line-of-sight direction
leading to a characteristic shape that seems to point at the observer. Somewhat misleadingly, these shapes were called \emph{Fingers-of-God}
\citep{1972MNRAS.156P...1J}. Lacking an analytical description, several empirical models exist that try to mimic this small-scale effect by smearing the density field along
the line-of-sight with the probability density function of velocities \citep{1972MNRAS.156P...1J, Hawkins:2002sg}. The models mainly differ in the assumption of how these
velocities are distributed; commonly, it is expected that they acquire a Maxwellian distribution, which gives rise to a multiplicative damping term for the power spectrum of
the form 
\begin{align} 
  F_\text{FoG}(k,\mu) = \left(1+\frac{k^2\,\mu^2\,\sigma_p^2}{2}\right)^{-1}\,.
\end{align}
The pairwise velocity dispersion $\sigma_p$ is approximately of  order $\sigma_p \sim 400\,$km/s \citep{Hawkins:2002sg}. 
\begin{figure}
  \centering
  \resizebox{\columnwidth}{!}{\input{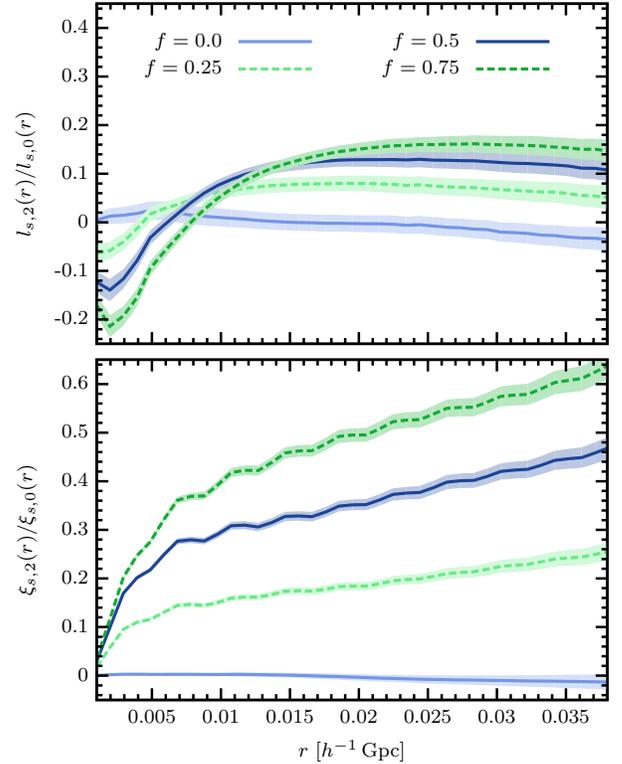}}
  \caption{Quadrupole-to-monopole ratio for the ALCF and two-point function calculated from a set of $50$ Zel'dovich density fields incorporating the Kaiser effect. The
    amplitude of the distortion is parametrised by $f$.}
  \label{fig:kaiser}
\end{figure}

The goal for the remainder of this section is to estimate how the ALCF is influenced by these kinematical redshift-space distortions. Hence, we take appropriate sets of mock
fields that simulate the Kaiser or Fingers-of-God effect and proceed as in the preceding section by computing the quadrupole-to-monopole ratio. The Kaiser effect can be easily
incorporated into the Zel'dovich density fields by boosting the Lagrangian displacement field in the line-of-sight direction by the factor $1+f$. This leads to distorted
density fields that seem to be derived from redshift-space. In Fig.~\ref{fig:kaiser}, we plot the quadrupole-to-monopole ratios for both the ALCF and the two-point function,
estimated from these fields where we leave $f$ as a free parameter which varies between $0.25$ and $0.75$. While the two-point function displays the anticipated behaviour,
i.e.~a squashing in the line-of-sight direction that becomes more prominent for increasing $f$, the ALCF exhibits a more complex quadrupole signal. On small scales, the
quadrupole is negative, indicating a stretching in the line-of-sight, before it changes sign at a crossover scale that seems to shift towards larger $r$ for higher $f$.
Thereafter, it remains positive, though the squashing is less pronounced than for the two-point function. Since the Kaiser effect shrinks the radial dimension of structures in
the density field, the enhancement of the transverse direction of the ALCF appears reasonable based on the discussion in Sec.~\ref{sec:ALCF.properties.angularScales}.

\begin{figure}
  \centering
  \resizebox{\columnwidth}{!}{\input{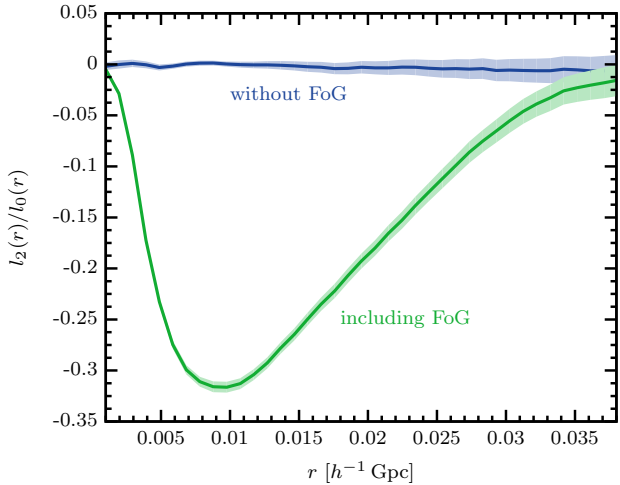}}
  \caption{ALCF quadrupole-to-monopole ratio for simple density fields that mimic the Fingers-of-God effect. Each field consists of $100$ spherical halos, each of which
    contains $1000$ particles that are Gaussian distributed about the centre with standard deviation $0.005\,L$. Their velocities are assigned via a Maxwellian distribution
    with velocity dispersion chosen such that the spread in position is $\sigma = 0.015\,L$. Results are based on $50$ realisations.}
  \label{fig:fog}
\end{figure}

Since the Zel'dovich approximation cannot account for the Fingers-of-God effect, we create a simplified setup which mimics these small-scale distortions. Firstly, we draw
random positions that are taken to be the centre points of halos. We assume that these halos have a spherical Gaussian density profile and accordingly draw particle positions
in each halo from a Gaussian distribution. Thereafter, we determine velocities from a Maxwellian distribution and displace all particles proportionally to their velocity along
the line-of-sight. Computing the quadrupole-to-monopole ratio gives the plot in Fig.~\ref{fig:fog}, showing a distinctive negative signal which confirms the stretching of
structures in the line-of-sight. In this simplified case, the Fingers-of-God effect establishes a clear preference of the line-of-sight, with no structures being aligned with
the transverse direction. However, as we have seen in Sec.~\ref{sec:ALCF.properties.numberdensities}, the signal of the ALCF may change drastically if this is the case, so it
would be interesting to see how the Fingers-of-God effect appears in a realistic N-body simulation that also contains filamentary structures in all directions.


\section{Conclusion}
\label{sec:conclusion}

In this paper, we presented an extension to the recently proposed line correlation measure \citep{linecorr} that enables the identification of anisotropies in a given density
field. The line correlation function is an estimator that relies purely on the phases of the density field and thus probes information not contained in ordinary two-point
statistics.

Starting from a generalisation of the originally defined line correlation function, we noticed that anisotropies tend to be smeared over the plane of radial and transverse
separations. For that reason, we introduced a novel mode cut-off scheme, which plays an integral part in the definition of the anisotropic line correlation function. We showed
that this new function produces better signal-to-noise ratios and, by using simple test fields as replacements for real observational data, we demonstrated that it extends the
properties of the original line correlation function in a consistent manner. In particular, we observed that the altered cut-off method allows for a clear distinction between
statistically isotropic and anisotropic fields and we showed that it is sensitive to angularly varying number densities and scales of structures.

In the second part, but still aiming primarily for a proof of concept, we considered observational effects, like the Alock-Paczy\'{n}ski, Kaiser and the Fingers-of-God
effect. We derived the relation between the assumed and true line correlation functions when redshifts and angles of galaxies are transformed with a wrong prior cosmology and
computed quadrupole-to-monopole ratios for all three effects, estimated from simple mock fields. We found that the line correlator is capable of detecting them, displaying
different systematics than the two-point function. While its sensitivity appears to be somewhat worse than that of the two-point function, this is still a promising result:
since the information probed by both measures is independent from each other, it is conceivable that employing both statistics can lead to tighter constraints on cosmological
parameters than, for instance, a combination of the two- and three-point function.

To investigate these prospects, a more detailed study of redshift-space distortions is necessary. It is inevitable to employ N-body simulations to analyse both the Kaiser and
Fingers-of-God effect in a realistic setup, and it would be interesting to see whether the crossover scale, which was observed in Fig.~\ref{fig:kaiser}, is indeed a persistent
feature. To understand this behaviour, additional analytical studies are desirable as well. A possible ansatz for such an investigation could be to reformulate the combination
of bispectrum and power spectra that appears in the perturbative expression of the line correlation function (see Eq.~(\ref{eq:ilcf-perturbative})) in terms of the
perturbative kernels in redshift-space which were derived in \cite{Scoccimarro1999}. Finally, it remains to be seen which parameters the line correlation function is able to
constrain and to what accuracy.

In addition to the independence of information probed by the two-point and line correlation functions, both measures have different systematics, which allows the breaking of
degeneracies in the parameter estimation. An interesting example is the logarithmic growth rate which describes the strenth of the Kaiser effect (see
Sec.~\ref{sec:RSD.kinematic}). Due to the fact that we are measuring galaxy positions, which are biased tracers of the underlying dark matter distribution, the two-point
function is actually only sensitive to the combination $\beta = f/b$, where $b$ is the linear bias parameter. On the other hand, the line correlation function is independent
of linear bias and hence constrains $f$ directly, meaning that a joint analysis might allow for the determination of $f$ and $b$ separately. We leave these exciting new
avenues for a future research project.


\section*{Acknowledgments}

We thank Danail Obreschkow for his useful feedback on the draft. We are also thankful for further comments from Diana~Battefeld and Ridwan Barbhuiyan. AE acknowledges the STFC
for his studentship funding.


\bibliographystyle{mn2e} 
\bibliography{refs}

\appendix

\section{Notes on the Implementation}
\label{sec:app.implementation}

In this appendix, we briefly comment on our numerical implementation of the line correlation function and show convergence tests. 

We adopt a standard numerical discretisation scheme: the density field is contained in a finite cubic box of sidelength $L$ on which we impose periodic boundary conditions.
The box is subdivided into an even number, $N$, of cells per side with the regular Cartesian grid spacing $\Delta x = L/N$. Accordingly, in Fourier space the box has a
sidelength $2\pi\,N/L$ with spacing $\Delta k = 2\pi/L$ and Nyquist frequency $k_{\text{Ny}} = \pi\,N/L$. Upon discretisation using this scheme, integrals in configuration and
Fourier space become
\begin{align}
  \int \D{^Dx}\,f(\B{x}) \hspace{1em} &\rightarrow \hspace{1em} \left(\frac{L}{N}\right)^D\,\sum_{\B{x}}\,f(\B{x})\,, \\
  \int \D{^Dk}\,f_{\B{k}} \hspace{1em} &\rightarrow \hspace{1em} \left(\frac{2\pi}{L}\right)^D\,\sum_{\B{k}}\,f_{\B{k}}\,.
\end{align}
Applying these rules to the prescription of the line correlation function in Eq.~(\ref{eq:ilcf-prescription}) we get
\begin{align}
  \label{eq:implementation-fourier}
  l(r) = \left(\frac{r}{L}\right)^{\frac{3D}{2}}\sum_{\substack{|\B{k}_1|,|\B{k}_2|,\\|\B{k}_1+\B{k}_2|\,\leq2\pi/r}}\,w_D(|\B{k}_1-\B{k}_2|r) \nonumber \\
  \times\,\frac{\delta_{\B{k}_1}\delta_{\B{k}_2}\delta_{-\B{k}_1-\B{k}_2}}{|\delta_{\B{k}_1}\delta_{\B{k}_2}\delta_{-\B{k}_1-\B{k}_2}|}\,.
\end{align}
Implementing this expression is straightforward, but computationally expensive due to the nested summations over $\B{k}_1$ and $\B{k}_2$. Hence, we use the equivalent of
Eq.~(\ref{eq:ilcf-prescription}) in configuration space, that is the spatially and rotationally averaged product of whitened density fields
\begin{align}
  l(r) = \left(\frac{r^D}{V}\right)^{\frac{3}{2}}& \int \frac{\D{^Dx}}{V}\,\left<\epsilon_{\theta}(\B{x})\right. \nonumber \\ 
    &\hspace{-2em}\left.\times\,\epsilon_{\theta}(\B{x}+{\cal R}\B{r})\,\epsilon_{\theta}(\B{x}-{\cal R}\B{r})\right>_{{\cal R}}\,,
\end{align}
where ${\cal R}$ denotes a rotation matrix and the mode cut-off is written as $\epsilon_{\theta}$ which stands for the convolution of $\epsilon(\B{x})$ with an appropriate
filter function $\Theta_{\B{r}}$. Discretising the equation above gives
\begin{align}
  l(r) =& \left(\frac{r}{L}\right)^{\frac{3D}{2}}\,N^{-D}\sum_{\B{x}}\left<\epsilon_{\theta}(\B{x})\right. \nonumber \\ 
  &\hspace{2em}\left.\times\,\epsilon_{\theta}(\B{x}+{\cal R}\B{r})\,\epsilon_{\theta}(\B{x}-{\cal R}\B{r}) \right>_{{\cal R}} \\
  \label{eq:implementation-config}
  =& \left(\frac{r}{L}\right)^{\frac{3D}{2}}\,\text{FFT}\left\{\left<\epsilon_{\theta}(\B{x})\right.\right. \nonumber \\
        &\hspace{2em}\left.\left.\times\,\epsilon_{\theta}(\B{x}+{\cal R}\B{r})\,\epsilon_{\theta}(\B{x}-{\cal R}\B{r})\right>_{{\cal R}}\right\}_{(\B{k} = 0)}\,,
\end{align}
where we replaced the summation over all $\B{x}$ by the Fast-Fourier-Transform (FFT) algorithm. Since we can quickly transform back and forth between Fourier and configuration
space using the FFT algorithm, it is advantageous to compute the convolution as well as the shift in position in Fourier space. There, shifting the field by the vector $\B{r}$
amounts to multiplication with the phase factor $\exp{(i\B{k}\cdot\B{r})}$, while minimising rounding errors that might arise from the gridding when evaluating the field at a
shifted position in configuration space. Hence, letting $\theta(\B{k},\B{r})$ either denote the original or modified cut-off in Eq.~(\ref{eq:ROT}), we have
\begin{align}
  \label{eq:implementation-conv+shift}
  \epsilon_{\theta}(\B{x}+\B{r}) = \text{FFT}^{-1}\left\{\epsilon_{\B{k}}\,\text{e}^{i\B{k}\cdot\B{r}}\,\theta(\B{k},\B{r})\right\}\,.
\end{align}
Eqs.~(\ref{eq:implementation-config}) and (\ref{eq:implementation-conv+shift}) form the basis of all numerical computations in this article. 

Let us consider a test case that allows us to compare with analytic results and study the convergence of the implementation with increasing $N$. We create density fields
consisting of $N_H$ spherical Gaussians with equal standard deviation $\sigma$. In this case (see Eqs.~(\ref{eq:halomodel-powerspec}) and (\ref{eq:halomodel-bispec})),
\begin{align}
  \label{eq:implementation-bispecpowerspec}
  \frac{B(\B{k}_1,\B{k}_2,\B{k}_3)}{\sqrt{P(|\B{k}_1|)\,P(|\B{k}_2|)\,P(|\B{k}_3|)}} = \sqrt{\frac{V}{(2\pi)^DN_H}}\,,
\end{align}
such that for $D=2$ and, after integrating out one azimuthal angle, Eq.~(\ref{eq:ilcf-perturbative}) gives
\begin{align}
  \label{eq:implementation-example.analytic}
  l(r) = 2\left(\frac{\sqrt{\pi}}{2}\right)^3\,&\bigg(\frac{r}{2\pi}\bigg)^3\sqrt{\frac{V}{N_H}}\,\int\displaylimits_0^{2\pi/r}\D{k_1}\,k_1
  \int\displaylimits_0^{2\pi/r}\D{k_2}\,k_2 \nonumber \\ &\times\,\int\displaylimits_{-1}^{\mu_{\text{cut}}}\frac{\D{\mu}}{\sqrt{1-\mu^2}}\,J_0(|\B{k}_1-\B{k}_2|\,r)\,,
\end{align}
where $\mu = \cos\sphericalangle(\B{k}_1,\B{k}_2)$ and its upper limit is given by $\mu_{\text{cut}} =
\text{min}\{1,\text{max}\{-1,[(2\pi/r)^2-k_1^2-k_2^2]/[2k_1k_2]\}\}$. Eq.~(\ref{eq:implementation-example.analytic}) can be integrated easily, for instance using a Monte-Carlo
approach, to yield the analytic answer. However, to compare with the numerical result we need to take care of a subtlety which arises in the whitening process: whenever the
amplitude of Fourier coefficients drops below a certain threshold value, their phase factors are not resolved properly any more; in order to avoid the introduction of
artifacts, we need to cut out these coefficients\footnote{This feature is due to the peculiar nature of the test fields considered here. Realistic density fields are unlikely
  to exhibit unresolved phase factors.}. We can apply the same rule as to the zero mode and set all modes with $|\delta_{\B{k}}| < 10^{-7}$ to zero. For our test fields there
exists an isotropic threshold scale $k_{\text{th}}$ so that all modes $k > k_{\text{th}}$ have unresolved phase factors. Hence, this effect can be dealt with in
Eq.~(\ref{eq:implementation-example.analytic}) by replacing the integration limits for $k_1$ and $k_2$ by $\text{min}\{k_{\text{th}},2\pi/r\}$ and similarly in
$\mu_{\text{cut}}$.
\begin{figure}
  \centering
  \resizebox{\columnwidth}{!}{\input{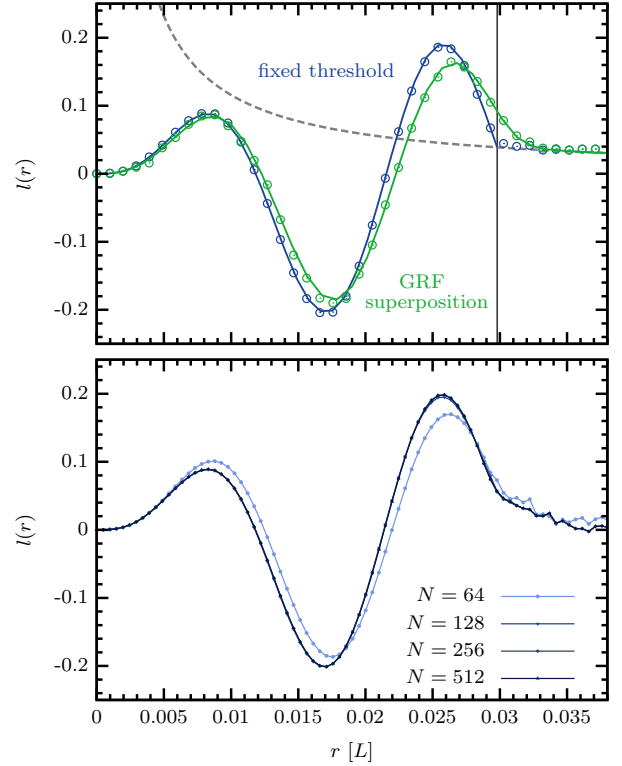}}
  \caption{Top panel: Comparison between analytics and numerics for a density field comprising $N_H = 100$ spherical Gaussian halos ($\sigma = 0.024\,L$). The dashed line is
    the result from Eq.~(\ref{eq:implementation-example.analytic}) while the blue and green solid lines either account for a fixed threshold scale or the superposition with a
    Gaussian random field with $P_{GRF} =\,$const.; the corresponding data points stem from the purely numerical implementation averaged over $50$ realisations. The vertical
    line indicates the threshold scale $r_{\text{th}} = 2\pi/k_{\text{th}}$. Bottom panel: Dependency on the number of grid points for a single realisation of the same density
    fields.}
  \label{fig:conv}
\end{figure}

Another way to deal with unresolved phase factors is the superposition of the density field with a Gaussian random field (GRF) of small amplitude to ensures that Fourier
coefficients with nearly vanishing amplitudes acquire random phases. This procedure guarantees that no artificial information compromises the signal: since the Gaussian random
field is uncorrelated to the Gaussian halos and assuming that it has zero mean, we only obtain a contribution to the overall power spectrum,
\begin{align}
  P(k) = P_G(k) + P_{GRF}(k)\,.
\end{align}
Doing so complicates Eqs.~(\ref{eq:implementation-bispecpowerspec}) and (\ref{eq:implementation-example.analytic}), but for known $P_{GRF}(k)$, it is still possible to find the
analytic answer via Monte-Carlo integration.  

The upper panel of Fig.~\ref{fig:conv} shows a comparison between numerical and analytic results. We see that to the right of the vertical line, which indicates the threshold
scale $r_{\text{th}} = 2\pi/k_{\text{th}}$, the data points approach the dashed line, which is the analytic result based on
Eq.~(\ref{eq:implementation-example.analytic}). Below $r_{\text{th}}$, the effect of cutting out unresolved phase factors becomes important and produces the oscillatory
behaviour, which can be exactly reproduced for both methods if we account for the corresponding modifications in Eq.~(\ref{eq:implementation-example.analytic}). Thus, this
test case serves as a convincing consistency check between our numerical scheme and the perturbative expansion of the line correlation function.

Finally, the lower panel of Fig.~\ref{fig:conv} displays a convergence test for a varying number of grid points using a single realisation of the test fields above. The plot
clearly shows that our method converges quickly, indicating only small discrepencies for $N$ as low as $N=128$. All of our computations were carried out with $N=512$ grid
points.

\end{document}